\documentstyle[epsfig,12pt]{article}
\begin{document}
\title{ The theoretical predictions for the study of the
$a_0(980)$ and $f_0(980)$ mesons in the $\phi$ radiative decays.
 \footnote{Talk presented by V.V. Gubin}}
\author{ N.N. Achasov and V.V. Gubin \\
 S.L. Sobolev Institute for Mathematics\\
630090 Novosibirsk 90, Russia }




\maketitle



\begin{abstract}
The potentialities of the production of the  $a_0$ and $f_0$ mesons in the 
$\phi$ radiative decays  are considered.
\end{abstract}


The central problem of light hadron spectroscopy has been the problem
of the scalar $f_0(980)$ and $a_0(980)$ mesons. It is well known fact
that these
states possess peculiar properties from the naive quark ($q\bar q$)
model point of view, see, for example
 \cite{achasov-84,achasov-91,achasov-1991,achasov-inp}.
 To clarify the nature of these mesons a number
of models has been suggested.
It was shown that all their challenging properties
could be understood \cite{achasov-84,achasov-91,achasov-1991,achasov-inp}
in the framework of the four-quark  ($q^2\bar q^2$) MIT-bag model 
\cite{jaffe-77}
  with symbolic quark structure 
 $f_0(980)=s\bar s(u\bar u+d\bar d)/
\sqrt{2}$ and $a_0(980)=s\bar s(u\bar u-d\bar d)/\sqrt{2}$. Along with the
 $q^2\bar q^2$ nature of  $a_0(980)$ and $f_0(980)$ mesons the possibility of
 their being the  $K\bar K$ molecule is discussed \cite{weinstein-90}.
During the last few years it was established
\cite{achasov-89,molecule,neutral} that the radiative decays
 of the $\phi$ meson $\phi\rightarrow\gamma f_0\rightarrow\gamma\pi\pi$ and
$\phi\rightarrow\gamma a_0\rightarrow\gamma\eta\pi$
could be a good guideline in distinguishing  the $f_0$ and $a_0$ meson
models. The branching ratios are considerably different in the cases of
naive quark, four-quark or molecular models. As has been shown
\cite{achasov-89,molecule,neutral}, in the four quark model the 
branching ratio is
\begin{equation}
BR(\phi\to\gamma f_0(q^2\bar q^2)\to\gamma\pi\pi)\simeq
BR(\phi\to\gamma a_0(q^2\bar q^2)\to\gamma\pi\eta)\sim10^{-4},
\end{equation}
and in the $K\bar K$ molecule model it is 
\begin{equation}
BR(\phi\to\gamma f_0(K\bar K)\to\gamma\pi\pi)\simeq
BR(\phi\to\gamma a_0(K\bar K)\to\gamma\pi\eta)\sim10^{-5}.
\end{equation}
It is easy to note that in the case $f_0=s\bar s$ and
$a_0=(u\bar u-d\bar d)/\sqrt{2}$ ( so called $s\bar s$ model \cite{tornqvist1}) the
branching ratios 
$BR(\phi\to\gamma f_0\to\gamma\pi\pi)$ and
$BR(\phi\to\gamma a_0\to\gamma\pi\eta)$ are different by factor of ten,
which should be visible experimentally.


 In the case when $f_0=s\bar s$ the suppression by the OZI rule is absent
and the evaluation gives \cite{achasov-89,neutral}
\begin{eqnarray}
BR(\phi\to\gamma f_0(s\bar s)\to\gamma\pi\pi)\simeq5\cdot10^{-5}, 
\end{eqnarray}
whereas for $a_0=(u\bar u-d\bar d)/\sqrt{2}$ the decay $\phi\to\gamma a_0\to
\gamma\pi\eta$ is suppressed by the OZI rule  and  is dominated by the real 
$K^+K^-$ intermediate state breaking the OZI rule 
\cite{achasov-89,neutral}
\begin{eqnarray}
BR(\phi\to\gamma a_0(q\bar q)\to\gamma\pi\eta)\simeq(5\div8)\cdot10^{-6}.
\end{eqnarray}


Imposing the appropriate photon energy cuts $\omega<100$ MeV, one can
show \cite{neutral}  that the background reactions
$e^+e^-\to\rho(\omega)\to\pi^0\omega(\rho)\to\gamma\pi^0\pi^0$,
$e^+e^-\to\rho(\omega)\to\pi^0\omega(\rho)\to\gamma\pi^0\eta$ and
$e^+e^-\to\phi\to\pi^0\rho\to\gamma\pi^0\pi^0(\eta)$ are negligible 
in comparison with the scalar meson contribution
 $e^+e^-\to\phi\to\gamma f_0(a_0)\to\gamma\pi^0\pi^0(\eta)$
for  $BR(\phi\to\gamma f_0(a_0)\to\gamma\pi^0\pi^0(\eta))$
greater than $5\cdot10^{-6}(10^{-5})$.


Let us consider the reaction $e^+e^-\to\phi\to\gamma (f_0+\sigma)\to\gamma
\pi^0\pi^0$ with regard to the mixing of the $f_0$ and $\sigma$ mesons.
We consider the one-loop mechanism of the $R$ meson production, where
$R=f_0,\sigma$, through the charged kaon loop, $\phi\to K^+K^-\to\gamma R$,
see \cite{{achasov-89},{molecule},{neutral}}. 
The whole formalism in the frame of which we study this problem is 
discussed in \cite{neutral}. The parameters of the $f_0$ and $\sigma$
mesons we obtain from fitting the $\pi\pi$ scattering data, see 
\cite{neutral}.
 


In the four-quark model and $s\bar s$ model we consider
 the following parameters  to be   free:
the coupling constant of the $f_0$ meson to the $K\bar K$
channel $g_{f_0K^+K^-}$, the coupling constant of the $\sigma$ meson
to the $\pi\pi$ channel $g_{\sigma\pi\pi}$, the constant of the
$f_0-\sigma$ transition  $C_{f_0\sigma}$,
the ratio $R=g^2_{f_0K^+K^-}/g^2_{f_0\pi^+\pi^-}$, the phase $\theta$ of the
elastic background and the $\sigma$ meson mass. The mass of the $f_0$ meson
is restricted to the region $0.97<m_{f_0}<0.99$ GeV. Treating the  $\sigma$
meson as an ordinary two-quark state, we get
 $g_{\sigma K^+K^-}=\sqrt{\lambda}
 g_{\sigma\pi^+\pi^-}/2\simeq0.35g_{\sigma\pi^+\pi^-}$, where
$\lambda\simeq1/2$ takes into account suppression of the strange quark production.
 So the constant
$g_{\sigma K^+K^-}$ ( and   $g_{\sigma\eta\eta}$ )  is not essential in 
our fit. 


As for the reaction $e^+e^-\to\gamma\pi^0\eta$  the similar analysis
of the $\pi\eta$ scattering cannot be performed directly. But, our analysis
of the final state  interaction for the $f_0$ meson production show that
the situation does not changed radically, in any case in the region
$\omega<100$ MeV. Hence, one can hope that the final state interaction
in the  $e^+e^-\to\gamma a_0\to\gamma\pi\eta$ reaction will not strongly
affect the predictions in the region $\omega<100$ MeV. 
Based on the analysis of $\pi\pi$ scattering and using the relations 
between coupling constants  we predict the quantities of the
 $BR(\phi\to\gamma a_0\to\gamma\pi\eta)$ in the  $q^2\bar q^2$ model, 
 $K\bar K$ model and the $q\bar q$ model where  $f_0=s\bar s$ and
$a_0=(u\bar u-d\bar d)/\sqrt{2}$.


The fitting shows that in the four quark model
($g^2_{f_0K^+K^-}/4\pi\sim 1\ GeV^2$ ) a number of parameters 
describe well enough the $\pi\pi$ scattering in the region $0.7<m<1.8$ GeV,
see \cite{neutral}. 
We predict 
$BR(\phi\to\gamma (f_0+\sigma)\to\gamma\pi\pi)\sim10^{-4}$ and
$BR(\phi\to\gamma a_0\to\gamma\pi\eta)\sim10^{-4}$ in the $q^2\bar q^2$
model. 



In the model of the $K\bar K$ molecule we get
$BR(\phi\to\gamma (f_0+\sigma)\to\gamma\pi\pi)\sim10^{-5}$ and
$BR(\phi\to\gamma a_0\to\gamma\pi\eta)\sim10^{-5}$.


In the $q\bar q$ model the $f_0(a_0)$ meson is considered as a
point-like object, i.e. in the  $K\bar K$ loop, $\phi\to K^+K^-\to\gamma
f_0(a_0)$ and in the transitions caused by the $f_0-\sigma$ mixing we consider
both the real and the virtual intermediate states. This model is different from
$q^2\bar q^2$ model by the coupling constant which is  $g^2_{f_0K^+K^-}/4\pi
<0.5\ GeV^2$.  In this model  we obtain
$BR(\phi\to\gamma (f_0+\sigma)\to\gamma\pi\pi)\simeq5\cdot10^{-5}$ and
taking into account the imaginary part of the decay amplitude only,
which violates the OZI rule, we get
$BR(\phi\to\gamma a_0(q\bar q)\to\gamma\pi\eta)\simeq8\cdot10^{-6}$.


The experimental data from SND and CMD-2 detectors support the four quark 
nature of the $f_0$  and $ a_0$  mesons, see Fig.1 and Fig.2.
and also \cite{printed,cmd,snd,golubev}.
 The obtained parameters for $f_0$ meson
from SND detector are $m_{f_0}=971\pm6\pm5$ MeV, $g^2_{f_0K^+K^-}/4\pi=
2.1\pm^{0.88}_{0.56}\ GeV^2$, $R=4.1$ and  
$BR(\phi\to\pi^0\pi^0\gamma)=(1.14\pm0.1\pm0.12)
\cdot10^{-4}$, see the dashed line on Fig.1. 



As for reaction $e^+e^-\to\gamma\pi^ + \pi^-$, the  analysis shows that 
the study of this reaction is an interesting and rather complex problem.


The main problem is the large background process of final pions radiation.
The $f_0$ state in this reaction could be studied only by observing
the interference patterns in the total cross-section and in the photon
spectrum \cite{interference,interference2}. As it was shown in 
\cite{interference2},
since the Fermi-Watson theorem for the final state interaction  
due to the soft photons 
in the reaction $e^ + e^-\to\rho(s)\to\gamma\pi^ + \pi^-$ is not valid,
the phase of the amplitude $\gamma^*(s)\to\rho\to\gamma\pi\pi$ does not 
determined by the s-wave phase of $\pi\pi$ scattering. The analyses of 
the interference patterns in the reaction 
$e^+e^-\to\phi+\rho\to\gamma f_0+\gamma\pi^+\pi^-\to\gamma\pi^+\pi^-$
should be performed taking into account the  phase of the elastic
background of the $\pi\pi$ scattering, the  phase of the triangle diagram
$\phi\to K^+K^-\to\gamma f_0$ and the phase of the $f_0$--$\sigma$
complex  in the $\phi\to K^+K^-\to\gamma(f_0+\sigma)\to\gamma\pi\pi$ 
amplitude.
 The whole formalism for the description of these reactions
and the resulting pictures were stated in 
\cite{neutral,interference,interference2}.


\begin{figure} 
\centerline{\epsfig{file=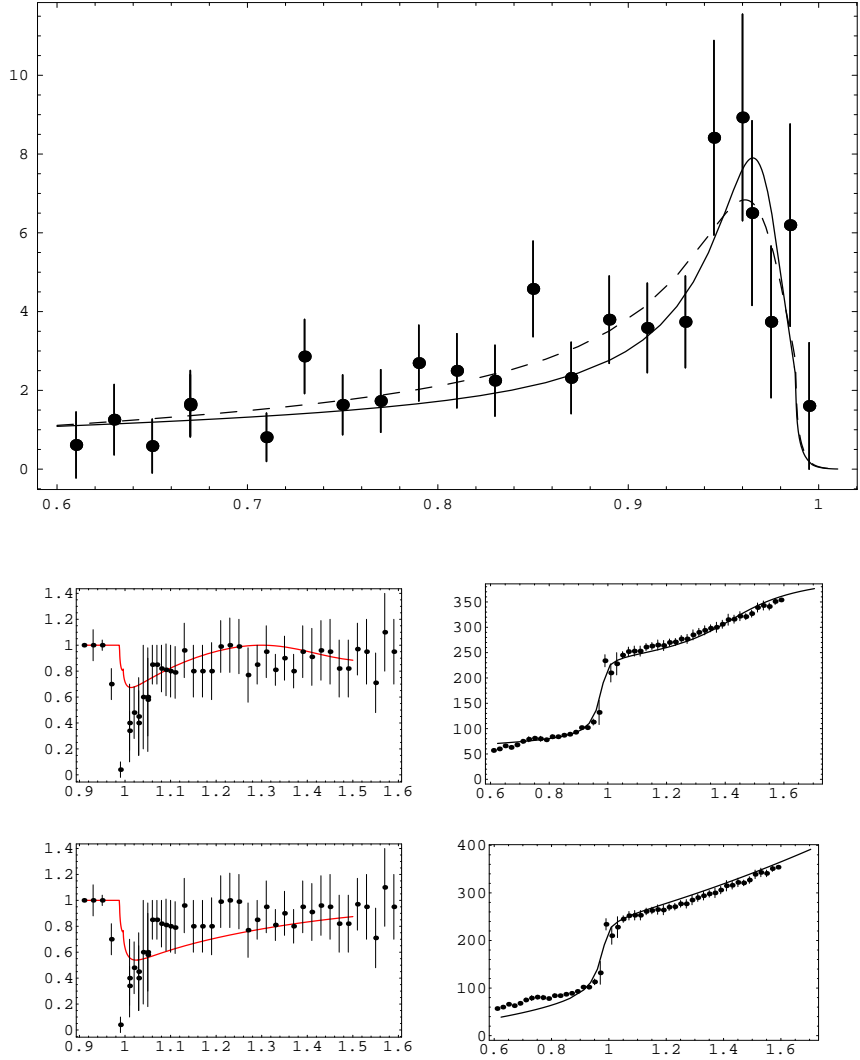,height=15cm,width=12cm}}
\vspace{10pt}
\caption{The simultaneous fit of the spectrum of the differential cross
section $d\sigma(e^+e^-\to\gamma(f_0+\sigma)\to\gamma\pi^0\pi^0)/d\omega$ 
with  mixing of the $f_0$ and $\sigma$ mesons (solid line) and of the 
$\pi\pi$ scattering data (first row).  The branching ratio for this 
fit is $BR(\phi\to\gamma\pi^0\pi^0)=2.8\cdot10^{-4}.$
The dashed line is the spectrum of the $f_0$ meson without mixing with
the $\sigma$ meson. The $\pi\pi$ scattering for this fit is in the second 
row.}
\end{figure}




\begin{figure} 
\centerline{\epsfig{file=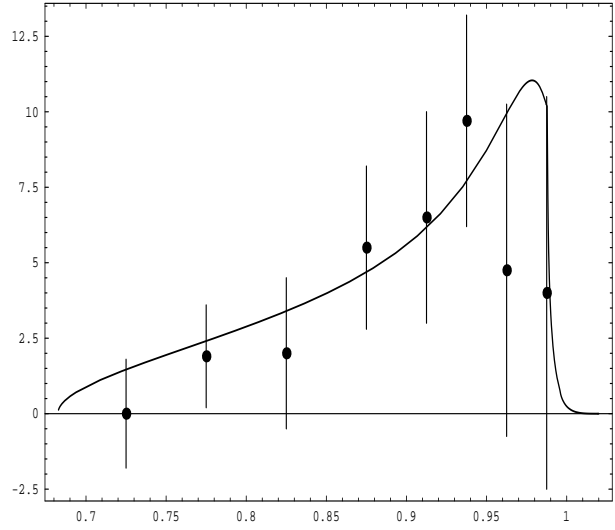,height=7cm,width=8cm}}
\vspace{10pt}
\caption{The fit of the spectrum of the differential cross
section $d\sigma(e^+e^-\to\gamma a_0\to\gamma\pi\eta)/d\omega$.
Parameters of the fit are $m_{a_0}=986\pm^{23}_{10}$ MeV,
$g_{a_0K^+K^-}^2/4\pi=1.5\pm0.5\ GeV^2$ and $BR(\phi\to\gamma\pi\eta)=
(0.83\pm0.23)\cdot10^{-4}.$}
\end{figure}





\end{document}